\documentclass[12pt]{article}
\usepackage{graphicx}

\begin{document}

\begin{titlepage}

\hfill{IPNO-DR/00-23}

\hfill{July 2000}

\vspace{2.5cm}

\begin{center}
{\Large\bf Vacuum fluctuations of $\bar{q}q$ and values\\
of low-energy constants\footnote{Work partly
supported by the EU, TMR-CT98-0169, EURODA$\Phi$NE network.}}

\vspace{1.5cm}

{\bf S. Descotes}\footnote{E-mail: {\tt descotes@ipno.in2p3.fr}}
and 
{\bf J. Stern}\footnote{E-mail: {\tt stern@ipno.in2p3.fr}}

\vspace{0.8cm}
{\it
IPN, Groupe de Physique Th\'eorique\\
Universit\'e de Paris-Sud\\
F-91406 Orsay Cedex, France}
\end{center}

\vspace{1.5cm}

\begin{abstract}
We discuss the influence of the vacuum fluctuations
of $\bar{q}q$ pairs on low-energy constants and condensates. 
The analysis of the Goldstone boson masses and decay constants shows that
the three-flavor condensate and some low-energy constants are very sensitive
to the value of $L_6$, which measures the Zweig-rule violation in
the scalar channel. A chiral sum rule based on experimental data in this
channel is used to constrain $L_6$, confirming a significant decrease
between the two- and the three-flavor condensates.
\end{abstract}

\vspace{1cm}

\vfill
\end{titlepage}

{\bf 1.}
Low-energy constants (LEC's) characteristic of the effective chiral
Lagrangian of QCD \cite{g-l}
are order parameters of spontaneous breaking of chiral
symmetry (SB$\chi$S) which encode informations about the chiral structure of
QCD vacuum. 
This is why their precise determination is of theoretical
importance. In QCD-like theories, the mechanism of SB$\chi$S is conveniently
described in terms of average properties of lowest Euclidean fermionic
modes \cite{dirac}. 
This framework singlets out the quark condensate $\langle\bar qq\rangle$ (i.e.
the average density of infrared fermionic modes) and the pion decay constant
$F_{\pi}$ (conductivity) as two prominent order parameters \cite{twoalt}. 

   Sofar, most determinations of low-energy constants
   \cite{g-l,largenc,abt}
   have operated under two
   assumptions: i) Quark condensate dominates the description of symmetry
   breaking effects, and in particular the expansion of Goldstone boson masses.
   ii) Quantum fluctuations are less important and the dynamics of SB$\chi$S
   approximately fits into the large-$N_c$ picture of QCD. A particular
   consequence of the latter assumption is the expectation that most order
   parameters will only marginally depend on the number $N_f$ of massless
   quarks: the two-flavor condensate
$\Sigma(2)=-\lim_{m_u,m_d \to 0}\langle\bar uu\rangle$ should, 
for instance, be almost
independent of the mass of the strange quark, even if the latter is set $0$
yielding the three-flavor condensate $\Sigma(3)$. There should be only one
condensate $\Sigma(2) \sim \Sigma(3)$, large and not affected by the vacuum
fluctuation of $\bar qq$ pairs. Simultaneously, certain LEC's like
$L_4(\mu)$  and $L_6(\mu)$ should be tiny, since they are 
large $N_c$-suppressed and they violate the Okubo-Zweig-Iizuka (OZI) rule.

        During the last years \cite{lattice,Bachir1,paramag,Bachir2}, 
	there has been a growing evidence taking its
 roots in the observed properties of the scalar $0^{++}$ channel against
 this simplified ``mean field approximation-type'' picture of SB$\chi$S. It has
 been argued that vacuum fluctuations of $\bar qq$ could affect some order
 parameters, that $\Sigma(3)$ can be about one half of $\Sigma(2)$
 \cite{Bachir1}, and
 that the low-energy constant $L_6$ could play the role of a third important order
 parameter describing fluctuations of the density of small fermionic modes
 \cite{paramag}.
 The purpose of this note is to report further arguments and evidences in
 favor of a particular role played by the constant  $L_6(\mu)$ and by vacuum
 fluctuations of $\bar qq$. The paper has two parts. First, we reconsider
 the standard determination of LEC's from Goldstone boson masses and decay
 constants as a function of the value of $L_6$. Next, we develop new
 arguments concerning a sum rule constraint \cite{Bachir1,Bachir2} on $L_6$ and on the 
 difference between the two-flavor
 and three-flavor quark condensates. 

{\bf 2.}
We first consider the pseudoscalar masses $M_{\pi}$, $M_K$, $M_{\eta}$ and the
corresponding decay constants (in the sequel we put $m_u=m_d=m$). We want to
demonstrate that the extraction of low energy parameters from these
observables is amazingly sensitive to the fine tuning of the OZI- and large
$N_c$-suppressed constant $L_6(\mu)$: in particular, a relatively tiny
increase of $L_6(\mu)$ implies a substantial drop of the three-flavor
condensate $\Sigma(3) = - \langle\bar uu\rangle_{m=m_s=0}$.  We start by
Eqs.~(10.7) of the classical Gasser-Leutwyler's paper \cite{g-l}
and we rewrite them as:
\begin{eqnarray}\label{pion}
F_{\pi}^2 M_{\pi}^2 &=& 2m \Sigma (3) + 2m(m_s+2m)Z + 4 m^2 A + 4 m^2 B_0^2 L
+ F_{\pi}^2 \delta_{\pi},\\
\label{kaon}
F_K^2 M_K^2 &=& (m_s+m) \Sigma(3) + (m_s+m)(m_s+2m)Z\\
&& \qquad +(m_s+m)^2 A + m(m_s+m)B_0^2 L + F_K^2 \delta_K.\nonumber
\end{eqnarray}
Here, $Z$ and $A$ are scale-independent constants containing the LEC's
$L_6(\mu)$  and $L_8(\mu)$,
\begin{eqnarray}\label{z}
Z &=& 32 B_0^2 \left[L_6(\mu) - \frac{1}{512 \pi^2}\left(\log\frac{M_K^2}{\mu^2}
+\frac{2}{9}\log\frac{M_{\eta}^2}{\mu^2}\right)\right],\\
\label{a}
A &=& 16 B_0^2 \left[ L_8(\mu) - \frac{1}{512 \pi^2}\left(
\log\frac{M_K^2}{\mu^2} +
\frac{2}{3}\log\frac{M_{\eta}^2}{\mu^2}\right)\right],
\end{eqnarray}
with $B_0=\Sigma(3)/F_0^2$ and $F_0 = \lim_{m,m_s\to 0} F_{\pi}$.
The remaining $O(p^4)$ chiral logarithms are contained in $L$:
$32 \pi^2 L = 3 \log(M_K^2/M_{\pi}^2) + \log(M_{\eta}^2/M_K^2)$,
$L=0.0253$.  There is a similar equation (not shown here) for
$F_{\eta}^2 M_{\eta}^2$,
which besides $\Sigma(3)$, $L_6(\mu)$ and $L_8(\mu)$
contains the LEC
$L_7$, as well as three equations for $F_P^2$ ($P$=$\pi$, $K$, $\eta$).
Notice that considering $F_P^2 M_P^2$ and $F_P^2$ as independent
observables allows one to separate from the start
the ``mass-type'' LEC's $L_6$, $L_7$,
$L_8$ from the constants $L_4$, $L_5$ which merely show up in the decay 
constants $F_P^2$.

In Eqs.~(\ref{pion}) and (\ref{kaon}),
all terms linear and quadratic in quark masses
 are explicitely shown and all remaining contributions starting by
 $O(m_\mathrm{quark}^3)$
 are gathered in the remainders $\delta_P$. We are going to
 imagine that the latter are given to us, allowing one to treat
 Eqs.~(\ref{pion})
 and (\ref{kaon}) as exact algebraic identities relating the three-flavor
 quark condensate $\Sigma(3)$ [expressed in physical units:
 $X(3)=2m\Sigma(3)/(F_{\pi}M_{\pi})^2$], the quark mass ratio $r=m_s/m$,
 and the LEC's $F_0$, $L_6(\mu)$ and $L_8(\mu)$. Although no expansion will
 be used, we want to investigate the consequences of the assumption
 $\delta_P \ll M_P^2$ on the LEC's. This might be interesting
 in connection with the recent discussion of convergence properties of
 three-flavor S$\chi$PT beyond $O(p^4)$ \cite{abt}.
 It should be stressed
 that even setting $\delta_P=0$, we do not follow the framework of S$\chi$PT
 at one-loop order \cite{g-l}: we do not assume that
 the condensate $\Sigma(3)$ dominates  in Eqs.~(\ref{pion}) and
 (\ref{kaon}), and consequently, we do not treat
 $1-X(3)$ as a small expansion parameter and do not replace in higher
 order terms $2mB_0$ by $M_{\pi}^2$. We do not follow
 G$\chi$PT either \cite{gchipt}, 
 since we do not treat $B_0$ as an expansion parameter:
 even with $\delta_P=0$, Eqs.~(\ref{pion}) and (\ref{kaon}) go
 beyond the tree level of G$\chi$PT since they include chiral logarithms.

 It is convenient to rewrite Eqs.~(\ref{pion})
 and (\ref{kaon}) as:
\begin{eqnarray}\label{cond}
 \frac{2m}{F_{\pi}^2 M_{\pi}^2} [\Sigma(3) + (2m+m_s)Z] &=& 1-
 \tilde\epsilon(r)- \frac{4m^2 B_0^2}{F_{\pi}^2 M_{\pi}^2} \frac{rL}{r-1}-\delta,\\
\label{eleight}
\frac{4m^2A}{F_{\pi}^2 M_{\pi}^2} &=& \tilde\epsilon(r) +      
\frac{4m^2B_0^2}{F_{\pi}^2 M_{\pi}^2} \frac{L}{r-1} + \delta',
\end{eqnarray}
where $\tilde\epsilon(r) = 2(\tilde{r}_2-r)/(r^2-1)$ with
$\tilde{r}_2= 2(F_KM_K)^2/(F_{\pi}M_{\pi})^2 -1$.
$\delta$ and $\delta'$ are simple linear combinations of $\delta_{\pi}$
and $\delta_K$:
\begin{eqnarray}\label{delta}
\delta &=& \frac{r+1}{r-1}\frac{\delta_{\pi}}{M_{\pi}^2} -
\left(\tilde\epsilon+\frac{2}{r-1}\right)\frac{\delta_K}{M_K^2}\\
\delta' &=& \frac{2}{r-1}\frac{\delta_{\pi}}{M_{\pi}^2}-
\left(\tilde\epsilon + \frac{2}{r-1}\right)\frac{\delta_K}{M_K^2}.
\end{eqnarray}
For large $r$, one expects
$\delta' \ll \delta \sim \delta_\pi/M_\pi^2$.
As in Ref.~\cite{abt}, we consider as input parameters 
$F_0$
[or equivalently $L_4(\mu)$], 
the large $N_c$-suppressed constant $L_6(\mu)$
and the quark mass ratio $r=m_s/m$. Eq.~(\ref{cond}) then yields the
following
non-perturbative formula for the three-flavor GOR ratio $X(3)$:
\begin{equation}\label{GOR}
X(3) = \frac
{2}{1+[1+\kappa(1-\tilde\epsilon-\delta)]^{1/2}}(1-\tilde\epsilon-\delta),
\end{equation}
where $\kappa$ contains the constant $L_6(\mu)$:
\begin{eqnarray}\label{kappa}
\kappa &=& 64 (r+2) \left(\frac{F_{\pi}M_{\pi}}{F_0^2}\right)^2 
\Bigg\{ L_6(\mu)\\
&&\qquad -\frac{1}{256\pi^2}\left(\log\frac{M_K}{\mu} +
\frac{2}{9}\log\frac{M_{\eta}}{\mu}\right) 
+ \frac{rL}{16(r-1)(r+2)}\Bigg\}.\nonumber
\end{eqnarray}

Eq.~(\ref{GOR}) is an exact identity which is useful to the extent that the
remainder $\delta$ in Eq.~(\ref{delta}) is small, i.e. if the expansion of
QCD correlation functions in powers of the quark masses $m_u$, $m_d$, $m_s$
globally converges. For instance, the latter requirement means that
$\delta_P\ll M_P^2$ in Eqs.~(\ref{pion}) and (\ref{kaon}), but not
necessarily that the linear (condensate) term dominates. 

\begin{figure}[t]
\begin{center}
\includegraphics[width=11cm]{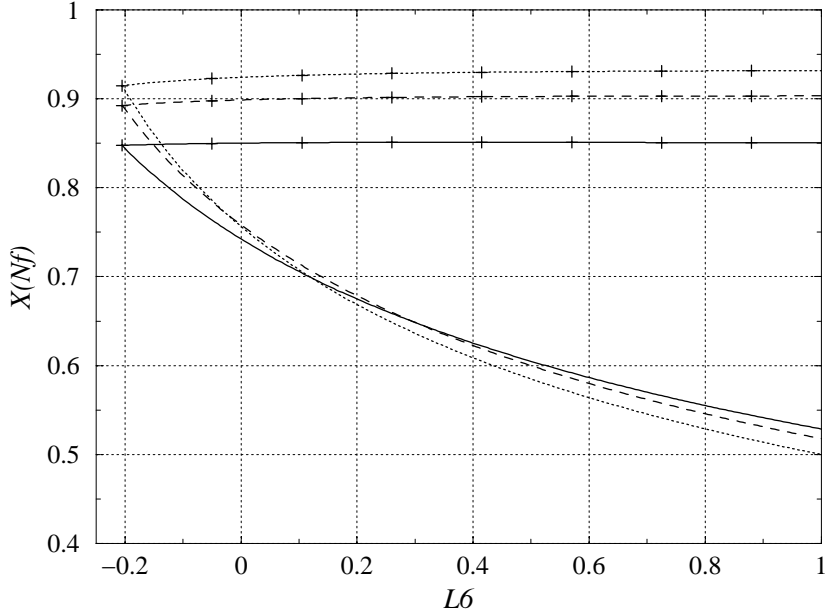}
\caption{$X(2)$ (lines with crosses) and $X(3)$ (lines with no symbol)
as functions of $L_6(M_\rho)\cdot 10^3$ and $r=m_s/m$ (solid: $r=20$, dashed: $r=25$, dotted:
$r=30$) for $F_0$=85 MeV.}
\label{x2x3}
\end{center}
\end{figure}

The parameter $\kappa$ describes quantum fluctuations of the condensate and,
indeed, $\kappa = 0(1/N_c)$. On the other hand, $\kappa$ is actually small
only if $L_6(\mu)$ is properly chosen: $\kappa=0$ for $ 10^3 L_6 = - 0.26$
at the scale $\mu=M_{\rho}$, which is close to the value advocated in
classical S$\chi$PT analysis \cite{daphne}.
In this case, Eq.~(\ref{GOR}) predicts $X(3)$
close to 1 unless the quark mass ratio $r$ substantially decreases and
$\tilde\epsilon \to 1$. This effect is well known in G$\chi$PT \cite{gchipt}.
Quantum fluctuations
can change this picture: the number in front of the curly bracket in
Eq.~(\ref{kappa}) is huge ($\sim 5340$ for $r=26$ and $F_0$ = 85 MeV) and,
consequently, even a rather small positive value of $L_6(M_{\rho})$ can imply
a substantial suppression of $X(3)$, independently of the value of $r=m_s/m$.
The effect is represented on Fig.~\ref{x2x3},
where $X(3)$ is plotted as a function
of $L_6(M_{\rho})$ for $r=20$, $r=25$, $r=30$ and
$F_0$ = 85 MeV. For smaller $F_0$, the decrease of $X(3)$ is even
slightly steeper.

Once $X(3)$ is determined one can use Eq.~(\ref{eleight}) in order to
infer the constant $L_8(\mu)$. The latter depends on $L_6$ merely through
$X(3)$: the constant $A$ defined in Eq.~(\ref{a}) contains
$B_0^2 L_8$, is given by
$r=m_s/m$ and is very little sensitive to $L_6$. Using the additional
equations for $F_{\eta}^2 M_{\eta}^2$ and for $F_{\pi}^2$, $F_K^2$,
$F_{\eta}^2$,
one can obtain $L_7$, $L_4$, $L_5$. The results are displayed in
Table~\ref{LECs}
as a function of $L_6$ for two values of $F_0$ and $r=25$.

\begin{table}[t]
\begin{center}
\begin{tabular}{|r@{.}l|r@{.}lr@{.}lr@{.}lr@{.}l|r@{.}lr@{.}lr@{.}lr@{.}l|}
\cline{3-18}
\multicolumn{2}{c}{}  &   \multicolumn{8}{|c}{$F_0$=75 MeV}  &
\multicolumn{8}{|c|}{$F_0$=85 MeV}\\
\hline
\multicolumn{2}{|c}{$L_6$} & \multicolumn{2}{|c}{4} & \multicolumn{2}{c}{5} & 
\multicolumn{2}{c}{7} & \multicolumn{2}{c}{8} & \multicolumn{2}{|c}{4} & 
\multicolumn{2}{c}{5} & \multicolumn{2}{c}{7} & \multicolumn{2}{c|}{8}\\
\hline
-0&2  &
-0&013 & 1&174 &-0&325 & 0&586 &
-0&282 & 1&603 & -0&503 & 0&994
\\
0&    & 
0&108 & 1&616 &-0&509 & 1&008 & 
 -0&242 & 1&993 & -0&699 & 1&440
\\
0&2   &
0&200 & 1&953 &-0&678 & 1&391 &
-0&210 & 2&306 & -0&880 & 1&848
\\
0&4   &
0&277 & 2&234 &-0&837 & 1&751 & 
-0&184 & 2&572 & -1&050 & 2&232
\\
1&    &
0&463 & 2&915 &-1&291 & 2&774 &
-0&116 & 3&228 & -1&532 & 3&314
\\
\hline
\end{tabular}
\caption{LEC's $L_i(M_\rho)\cdot 10^3$ as functions of
$L_6(M_\rho)\cdot 10^3$ and $F_0$, for $r=25$.}
\label{LECs}
\end{center}
\end{table}

{\bf 3.}
It has been shown elsewhere \cite{paramag} that vacuum fluctuations of
$\bar qq$ imply an enhancement of the two-flavor condensate
$\Sigma(2)=-\lim_{m \to 0}\langle\bar uu\rangle|_{m_s\ \mathrm{fixed}}$ 
with respect to
$\Sigma(3)$, and consequently: $X(2) > X(3)$.
 The two-flavor condensate can be obtained as a limit:
\begin{equation}\label{sigma2}
\Sigma(2) = \lim_{m\to 0}\frac{(F_{\pi}M_{\pi})^2}{2m} = \Sigma(3) + 
m_s Z|_{m=0} + \delta \Sigma(2)
\end{equation}
keeping $m_s$ fixed. We have $Z|_{m=0}= Z + B_0^2 \Delta Z$, with
$\Delta Z=[\log(M_K^2/\bar M_K^2)+ 2/9 \cdot \log(M_{\eta}^2/\bar M_{\eta}^2)]/(16\pi^2)$
and $\bar M_P^2 = \lim_{m\to 0} M_P^2$. 
$\Delta Z$ can be obtained as a corollary of the previous analysis
of masses and decay constants, and it has a very little effect.
It has to be compared to the logarithmic piece of $Z$ in Eq.~(\ref{z}), taken
at a typical scale $\mu\sim M_\rho$. $\Delta Z$ is less than the tenth of
this logarithmic term.

Eliminating $Z$ from Eqs.~(\ref{cond}) and (\ref{sigma2}),
one arrives at the expression for the two-flavor GOR ratio
$X(2) = 2m\Sigma(2)/(F_{\pi}M_{\pi})^2$:
\begin{eqnarray}\label{xtwo}
X(2)&=& [1-\tilde\epsilon]\frac{r}{r+2} + \frac{2}{r+2}X(3)\\
&&\qquad -\frac{(F_{\pi}M_{\pi})^2}{2 F_0^4}X(3)^2 
\left[\frac{2r^2}{(r-1)(r+2)}L
-r\Delta Z\right] + \delta_X.\nonumber
\end{eqnarray}
In the expression for $\delta_X$, the remainders $\delta_P/M_P^2$ are
suppressed by $m/m_s$: for $r>20$ one expects
$|\delta_X| \sim |\delta'|$.
The dependence of $X(2)$ on $L_6$ is
entirely hidden in $X(3)$ and consequently it is rather marginal,
as can be seen from Fig.~\ref{x2x3},
where $X(2)$ is shown together with $X(3)$ for
$r=20$, 25, 30. ($X(2)$ however remains strongly correlated with $r=m_s/m$,
in particular for smaller values of $r$.) Fig.~\ref{x2x3} clearly exhibits
the lower bound $L_6(M_{\rho})> -0.21\cdot 10^{-3}$, as a consequence of the
paramagnetic inequality $X(3)<X(2)$. A similar inequality holds for the
decay constants $F(3) < F(2)$, where $F(N_f)$ denotes $F_{\pi}$ in the
limit of first $N_f$ massless quarks. This inequality implies a lower bound for
the large $N_c$-suppressed constant $L_4$. Equivalently, it can be
reexpressed as an upper bound for $F(3) = F_0$: $F_0 < F(2) \sim  87$ MeV.
$F_0$ substantially smaller than this upper bound corresponds to a larger
positive value of $L_4$. For instance, former studies of OZI-violation effects
in the slopes of scalar form factors \cite{Donoghue}
suggest a rather strong effect
of $L_4$ (see also Ref.~\cite{Meissner})
corresponding to the values $F_0$ as low as 71 MeV.

{\bf 4.}
The standard values of LEC's (see e.g. Ref.~\cite{daphne}), including the presumed
OZI-rule suppression of $L_6(\mu)$ and $L_4(\mu)$ are, of course, compatible
with the previous analysis. This is illustrated in the second half of the
first line of Table \ref{LECs} together with corresponding values
$X(2) \sim X(3) \sim 0.9$. The analysis of Goldstone boson masses and decay
constants alone does not allow to conclude that vacuum fluctuations of
$\bar qq$ actually produce an important effect. It however reveals an extreme
sensitivity of the resulting LEC's to the input value of $L_6(\mu)$: a modest
shift of the latter towards small positive values produces an important
splitting of $X(2)$ and $X(3)$ and an important increase of $L_8(\mu)$ and
$L_5(\mu)$. An insight on the actual effect of vacuum fluctuations of $\bar qq$
can be obtained from sum rules for the connected part of the large 
$N_c$-suppressed two-point function $\langle\bar uu(x)\bar ss(0)\rangle^c$ first analyzed in
Ref.~\cite{Bachir1}. In the remaining part of this work,
we reexamine the constraints imposed
by this sum rule on $L_6(\mu)$ and $X(3)$ avoiding to use S$\chi$PT and, in
particular, to treat $1-X(3)$ as a small expansion parameter. Our approach    
will be based on the the preceding analysis of masses and decay constants.
In addition, we shall use Operator Product Expansion
(OPE) of QCD to better control the high-energy
contribution to the sum rule.

{\bf 5.}  Let us consider the 
correlator:
\begin{equation}
\Pi(p^2)=i\frac{m m_s}{M^2_\pi M^2_K}
   \lim_{m\to 0} \int d^4x \ e^{ip\cdot x}\ 
   \langle 0|T\{\bar{u}u(x)\ \bar{s}s(0)\}|0\rangle^c,
\end{equation}
which is invariant under the QCD renormalization group and violates the
OZI rule in the $0^{++}$ sector.
For $m_s\neq 0$, $\Pi$ is an order parameter for
$\mathrm{SU}_L(2)\times\mathrm{SU}_R(2)$, related to the
derivative of $\Sigma(2)$ with respect to $m_s$: 
$m m_s\partial \Sigma(2)/\partial m_s=M_\pi^2 M_K^2 \Pi(0)$.
Differentiating Eq.~(\ref{sigma2}) with respect to $m_s$ and using Eq.~(\ref{z})
to compute $\partial Z/\partial m_s$, we obtain
a relation between $\Pi(0)$ and $Z$ \cite{paramag}:
\begin{eqnarray}
&&X(2)-X(3)=2\frac{mm_s}{F_\pi^2M_\pi^2}
  \left.Z \right|_{m=0}+
  \frac{2m}{F_\pi^2 M_\pi^2} \lim_{m\to 0}\frac{F_\pi^2\delta_\pi}{2m}
=2\frac{M_K^2}{F_\pi^2}\Pi(0)\label{diffgor}\\
&&\qquad+\frac{r[X(3)]^2}{32\pi^2}\frac{F_\pi^2M_\pi^2}{F_0^4}
    \left[\bar\lambda_K+\frac{2}{9}\bar\lambda_\eta\right]
  +\frac{2m}{F_\pi^2M_\pi^2}\left[1-m_s\frac{\partial}{\partial m_s}\right]
   \lim_{m\to 0}\frac{F_\pi^2\delta_\pi}{2m},\nonumber
\end{eqnarray}
where $\bar\lambda_P=m_s\cdot\partial[\log \bar{M}_P^2]/\partial m_s$.
We want to bring new information by estimating $\Pi(0)$ from a QCD sum rule.
The behaviour of $\Pi$ at large
momentum is given by OPE.
It will involve operators transforming as
$(\bar{u}u)(\bar{s}s)$ under the chiral group. In the limit
$m\to 0$, the lowest-dimension operator is $m_s\langle\bar{u}u\rangle$.
At the leading order in $\alpha_s$, the Wilson coefficient of this
operator can be computed from two-loop Feynman diagrams, one loop
stemming from the
$\bar{s}s$ source and another one from $\bar{u}u$ (the latter is open,
due to the insertion of the quark condensate). These quark loops are connected by 
two gluon lines.
Hence, we deal with genuine two-loop integrals, formally similar to those involved in
self-energy diagrams. The asymptotic expansion technique
allows to simplify the computation of the
large-momentum behaviour of such integrals,
by performing various Taylor expansions of the propagators \cite{largemoment}. This tedious
calculation ends up with the lowest-dimension term in the Operator Product Expansion
for $\Pi$:
\begin{equation}
\Pi(p^2)=-\frac{18[1-2\zeta(3)]}{P^2}\left(\frac{\alpha_s}{\pi}\right)^2
      \frac{m_s^2 m\langle\bar{u}u\rangle}{M_\pi^2M_K^2}+\ldots \label{asympt}
\end{equation}
where $P^2=-p^2$. Taking into account the running of $\alpha_s$ and of
the masses, we see that $\Pi$ vanishes faster than $1/p^2$ and is
superconvergent \cite{Bachir1}.

We can write a sum rule for this correlator at zero momentum:
\begin{eqnarray}
\Pi(0)&=&
  \frac{1}{\pi}\int_0^{s_1} ds\ \mathrm{Im}\ \Pi(s) 
     \ \frac{1}{s}\left(1-\frac{s}{s_0}\right)\label{regsom}\\
&&+\frac{1}{\pi}\int_{s_1}^{s_0} ds\ \mathrm{Im}\ \Pi(s)
     \ \frac{1}{s}\left(1-\frac{s}{s_0}\right)
  + \frac{1}{2i\pi}\int_{|s|=s_0} ds\ \Pi(s) 
     \ \frac{1}{s}\left(1-\frac{s}{s_0}\right).\nonumber
\end{eqnarray}
The integral is made of 3 pieces. Along the real axis, from 0 to
$\sqrt{s_1}\sim 1.2\ \mathrm{GeV}$, we compute explicitly the spectral
function $\mathrm{Im}\ \Pi(s)$ by solving two-channel coupled
Omn\`es-Muskhelishvili equations \cite{Bachir1,Bachir2,Donoghue}
for various $T$-matrix models \cite{Au,Loiseau,Oset}. The second
integral along the real axis
($\sqrt{s_1}\leq \sqrt{s}\leq \sqrt{s_0}\sim 1.6\ \mathrm{GeV}$) is roughly
estimated using a second sum rule. The last integration (around the complex
circle) is performed supposing that duality applies in the scalar channel
above $s_0$. 

{\bf 6.}
For $\sqrt{s}\leq \sqrt{s_1}\sim 1.2\ \mathrm{GeV}$, $\pi\pi$ and $K\bar{K}$
are expected to dominate the spectral function. No apparent
$4\pi$- and $\eta\eta$-contributions are observed in this energy range, and
hence we will neglect these channels. We will follow closely the analysis of
Refs.~\cite{Bachir1,Bachir2}. The spectral
function can be expressed in terms of the pion and kaon scalar
form-factors:
\begin{equation}
\vec{F}(s)=
  \left(
  \begin{array}{c}
  \langle 0|\bar{u}u|\pi\pi\rangle\\
  \langle 0|\bar{u}u|K\bar{K}\rangle
  \end{array}
  \right), \qquad
\vec{G}(s)=
  \left(
  \begin{array}{c}
  \langle 0|\bar{s}s|\pi\pi\rangle\\
  \langle 0|\bar{s}s|K\bar{K}\rangle
  \end{array}
  \right). 
\end{equation}
$\vec{F}$ and $\vec{G}$ 
satisfy (separately) a set of coupled Mushkelishvili-Omn\`es integral
equations:
\begin{equation}
F_i(s)=\frac{1}{\pi}\sum_{i=1}^n\int_{4M_\pi^2}^\infty ds' \frac{1}{s'-s}
   T_{ij}^*(s') \sqrt\frac{s'-4M_j^2}{s'} \theta(s'-4M_j^2) F_j(s'),
   \label{omeq}
\end{equation}
provided that the $T$-matrix behaves correctly when $s\to\infty$.
Obviously, new channels open when the energy increases, which invalidates the two-channel
approximation. However, we want $\vec{F}$ and $\vec{G}$ for $s\leq s_1$, and
the shape of the spectral function at much higher energies is not important.
For our purposes, we can therefore impose on our $T$-matrix model
any convenient large-energy behaviour.
If $\Delta(s)$ denotes the sum of the diagonal phase shifts, 
the condition $\Delta(\infty)-\Delta(4M_\pi^2)= 2\pi$ insures the existence and the unicity
of the solution for the linear equation (\ref{omeq}), once the values of the form factors at zero
are fixed \cite{Bachir1}. Any solution can be projected on a basis of two solutions $A$ and
$B$ such that $\vec{A}(0)={1\choose 0}$
and $\vec{B}(0)={0\choose 1}$ \cite{Bachir2}:
\begin{equation}
\vec{F}(s)=F_1(0)\vec{A}(s)+F_2(0)\vec{B}(s),\qquad
\vec{G}(s)=G_1(0)\vec{A}(s)+G_2(0)\vec{B}(s). \label{formfact}
\end{equation}
The values of the scalar form factors at zero, $\vec{F}(0)$ and $\vec{G}(0)$, are
related to the derivatives of the pion and kaon masses
with respect to $m$ and $m_s$:
\begin{equation}
\gamma_P
   =\frac{m}{M^2_P}\left(\frac{\partial M_P^2}{\partial m}\right)_{m=0},
 \qquad
\lambda_P
   =\frac{m_s}{M^2_P}\left(\frac{\partial M_P^2}{\partial m_s}\right)_{m=0},
 \qquad P=\pi,K. \label{deriv}
\end{equation}
In particular, $G_1(0)=(\partial M_\pi^2/\partial m_s)_{m=0}=0$. The
spectral function is the sum of two terms:
\begin{eqnarray}
\mathrm{Im}\ \Pi(s)&=&
  \gamma_\pi\lambda_K
   \left[\frac{\sqrt{3}}{32\pi}
      \sum_{i=1,2} \sqrt{\frac{s-4M_i^2}{s}} A_i(s) B^*_i(s)
      \theta(s-4M_i^2)\right]\label{foncspec}\\
&& +\gamma_K\lambda_K\frac{M_K^2}{M_\pi^2}
      \left[\frac{1}{16\pi}\sum_{i=1,2} \sqrt{\frac{s-4M_i^2}{s}} B_i(s) B^*_i(s)
      \theta(s-4M_i^2)\right]. \nonumber
\end{eqnarray}
The first integral in Eq.~(\ref{regsom}) is therefore
of the form
$\gamma_\pi\lambda_K \mathcal{I}_{AB}(s_0,s_1)+\gamma_K\lambda_K\mathcal{I}_{BB}(s_0,s_1)$.
$\mathcal{I}_{AB}$ and $\mathcal{I}_{BB}$ depend only on $s_0$, $s_1$, and on the chosen
$T$-matrix model \cite{Au,Loiseau,Oset}.
On the other hand, $\gamma_\pi$, $\gamma_K$ et $\lambda_K$
can be expressed through the expansion of pseudoscalar masses and decay
constants in powers of the quark masses. The main difference with respect to
Refs.~\cite{Bachir1,Bachir2} lies here in the normalization of the integrals
stemming from $\vec{F}(0)$ and $\vec{G}(0)$. As already emphasized, we do
not follow the standard $O(p^4)$ analysis of the
logarithmic derivatives in Eq.~(\ref{deriv}) \cite{Bachir1}. Even if the
three-flavor condensate does not dominate, 
$\lambda_P$ and $\gamma_P$ can be determined
from equations for $F_P^2 M_P^2$ and $F_P^2$ similar to
Eqs.~(\ref{pion}) and (\ref{kaon}).

{\bf 7.} The contribution of the integral under $s_1$ is positive and
dominated by the
$f_0(980)$ peak. On the other hand, $\Pi$ is superconvergent 
and the integral of the spectral 
function from 0 to $\infty$ is zero. $\mathrm{Im}\ \Pi(s)$
should therefore be negative somewhere.
Let us suppose that it is the case for
$s_1\leq s \leq s_0$.\footnote{If the spectral function is partially
positive in this range, our assumption yields an estimate for the second integral
in Eq.~(\ref{diffgor}) which is lower than in reality. In this case,
we will underestimate $X(2)-X(3)$.} We can then
estimate the contribution of the intermediate region in Eq.~(\ref{diffgor}) by:
\begin{equation}
\frac{1}{s_0} \mathcal{J}' \leq 
-\frac{1}{\pi}\int_{s_1}^{s_0} ds\ \mathrm{Im}\ \Pi(s)
     \ \frac{1}{s}\left(1-\frac{s}{s_0}\right)
\leq \frac{1}{s_1} \mathcal{J}', \label{estimate}
\end{equation}
where the integral $\mathcal{J}'$ can be estimated, using a second sum rule:
\begin{eqnarray}
\mathcal{J}'
  &=&-\frac{1}{\pi}\int_{s_1}^{s_0} ds\ \mathrm{Im}\ \Pi(s)
     \ \left(1-\frac{s}{s_0}\right)\\
  &=&\frac{1}{\pi}\int_0^{s_1} ds\ \mathrm{Im}\ \Pi(s) 
     \ \left(1-\frac{s}{s_0}\right)
+ \frac{1}{2i\pi}\int_{|s|=s_0}\!\!\! ds\ \Pi(s) 
     \ \left(1-\frac{s}{s_0}\right).\nonumber
\end{eqnarray}
The first integral on the right hand-side is known from the previously computed
spectral function (\ref{foncspec}), 
and the second one will be evaluated by replacing $\Pi$ by its
asymptotic behaviour according to OPE, cf. Eq.~(\ref{asympt}).

The third integral in Eq.~(\ref{regsom}) is estimated through the OPE
asymptotic behaviour of $\Pi$. The factor $(1-s/s_0)$ ensures that this
contribution is suppressed for $s$ in the vicinity of $s_0$. Notice that in
Eq.~(\ref{asympt}), $\langle\bar{u}u\rangle$
stands for the quark condensate with $N_f=2$ (we take the chiral
limit $m\to 0$, but $m_s$ remains at its physical value). 
Its contribution can be calculated using standard techniques,
relying on the expansion of the coupling constant along the circle
\cite{Braaten}:
\begin{eqnarray}
&&\frac{1}{2i\pi}\int_{|s|=s_0}\!\!\! ds\ \Pi(s) 
     \ \frac{1}{s}\left(1-\frac{s}{s_0}\right)
=9[1-2\zeta(3)]\\
&&\qquad \qquad \qquad \times
\frac{F_\pi^2}{M_K^2} X(2) \frac{m^2_s(s_0)}{s_0}
  a(s_0)^2[1-6.5\cdot a(s_0)+48.236\cdot a^2(s_0)],\nonumber
\end{eqnarray}
where $a=\alpha_s/\pi$.
The contribution is negative, but strongly suppressed by $\alpha^2_s$ and
$m_s^2/s_0$. Let us remember that, due to the probable importance of direct
instanton contributions \cite{directinst}, the duality in the $0^{++}$
channel is not expected to extend to as low energies as in other channels.

\begin{figure}[t]
\begin{center}
\includegraphics[width=11cm]{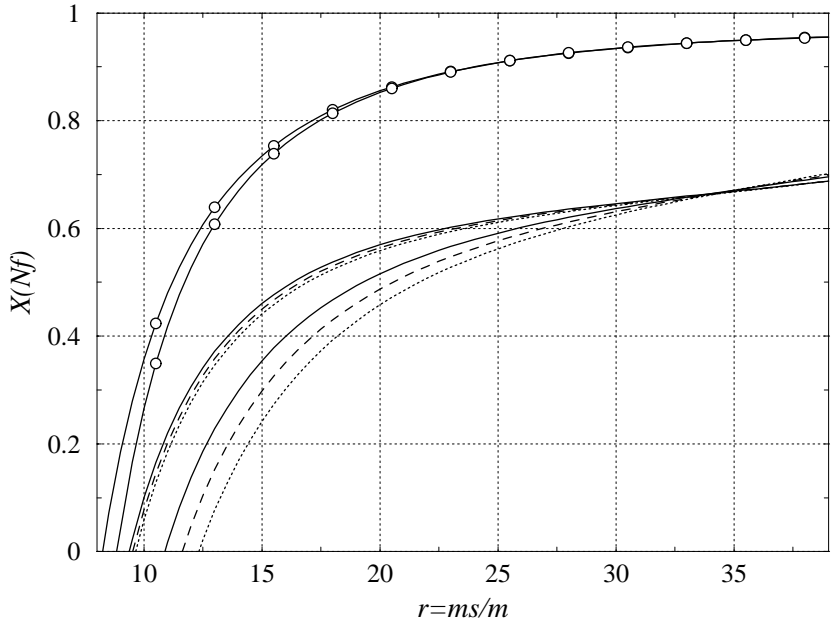}
\caption{Upper and lower bounds for $X(3)$ as functions of
$r=m_s/m$ for $F_0$=85 MeV,
with the T-matrix model of Ref.~\cite{Oset} for $s_1$=1.2 GeV and $s_0$=1.5
GeV (solid lines), 1.6 GeV (dashed lines) and 1.7 GeV (dotted lines). The
allowed range for $X(2)$ is plotted as a guide (line with an open circle).} 
\label{x2x3oset}
\end{center}
\end{figure}

{\bf 8.}
Hence, we can estimate $X(2)-X(3)$ in two different ways : the first one
is Eq.~(\ref{xtwo}), whereas the second one combines Eq.~(\ref{diffgor}) and
the sum rule Eq.~(\ref{regsom}). In both cases, $X(2)-X(3)$ can be expressed
as a function of observables
and of $[F_0,r,X(3)]$. This overdetermination can be considered as
a constraint fixing $X(3)$ as a function of $r$ and $F_0$. A typical result
is shown on Fig.~\ref{x2x3oset} for $F_0=85$ MeV.

\begin{figure}[t]
\begin{center}
\includegraphics[width=11cm]{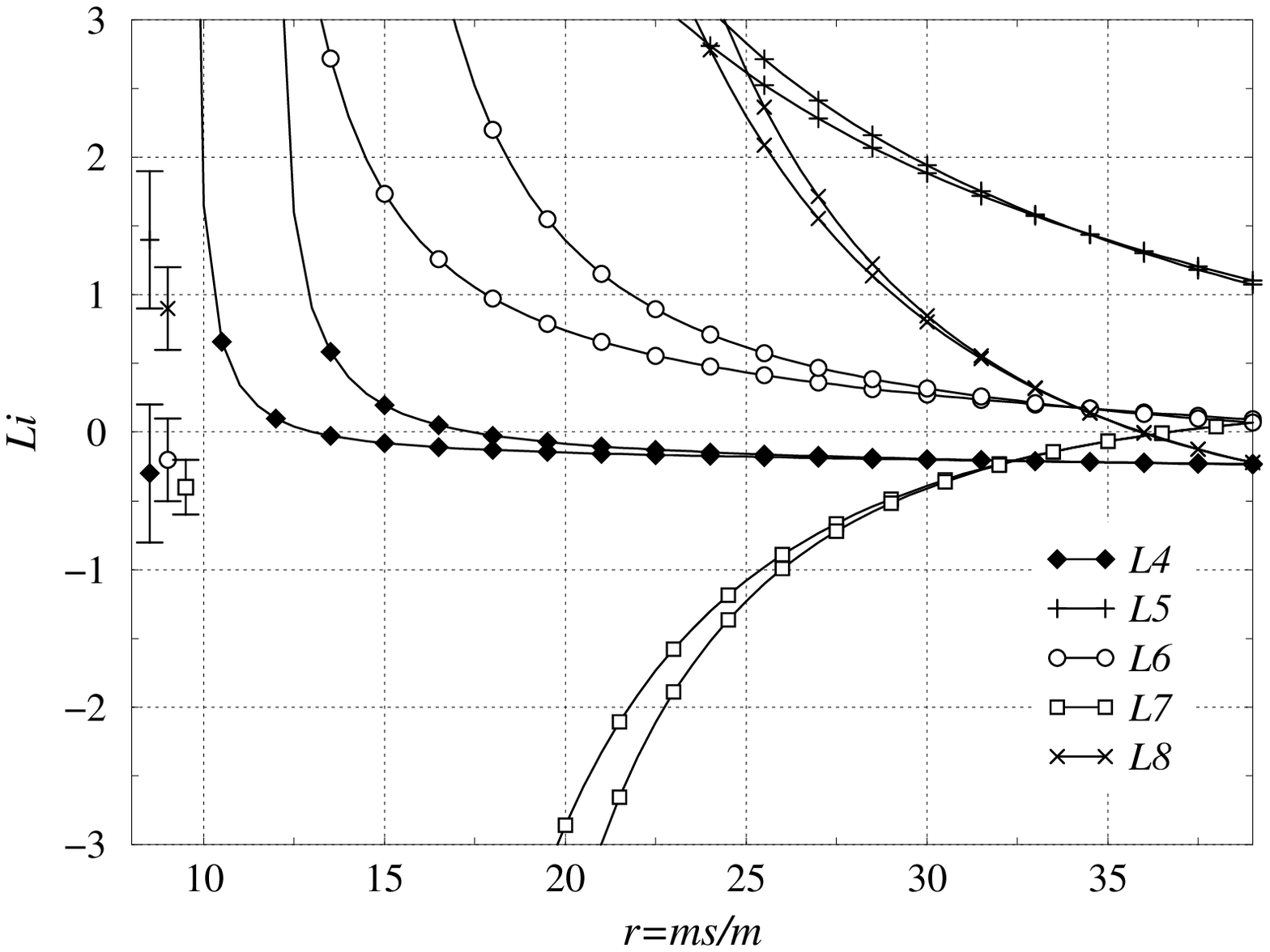}
\caption{Low-energy constants $L_{i=4\ldots 8}(M_\rho)\cdot 10^3$ as
functions of $r=m_s/m$ for $F_0$=85 MeV, with the T-matrix model of
Ref.~\cite{Oset} and $s_1$=1.2 GeV, $s_0$=1.6 GeV. The standard values
plotted on the left are taken from Ref.~\cite{daphne}.}
\label{lecplot}
\end{center}
\end{figure}

There are three sources of uncertainties in this analysis. {\it i)} First, there are
next-to-next-to-leading (NNL) remainders in the expressions for masses and
decay constants. The control of their effect is rather clear in
Eq.~(\ref{xtwo}) ($\delta_X\sim\delta'$), but somewhat less transparent in
Eq.~(\ref{diffgor}) and in the estimates of the logarithmic derivatives
$\lambda_P$ and $\gamma_P$. Within S$\chi$PT, authors of Ref.~\cite{abt}
have found that the two-loop effects on the $m_s$-dependence of $\Sigma(2)$
are likely to be small and of the same sign as the one-loop effects. A
similar conclusion is reached in Ref.~\cite{Bachir2}. NNL remainders are
assumed here to be small, and they are not shown in our results. 

{\it ii)}
Next, the evaluation of the sum rule Eq.~(\ref{regsom}) involves a rough
estimate of the integral between $s_1$ and $s_0$. Hence, for a given couple
$(F_0,r)$, the present evaluation of the sum rule will not provide a
single value of $X(3)$, but a whole range of acceptable values, depending in
addition on the separators $s_1<s_0$. It is seen on Fig.~\ref{x2x3oset} that
the upper bound for $X(3)$ remains stable for $\sqrt{s_0}>1.5$ GeV, whereas
the lower bound depends rather strongly on $s_0$. When $s_0$ increases, the
lower bound in Eq.~(\ref{estimate}) appears too weak to be saturated. A more
stringent lower bound would be welcome. 

{\it iii)}
The third source of uncertainty is the multi-channel $T$-matrix used as
input to build the spectral function in Eq.~(\ref{foncspec}) for $s<s_1$. Three
different $T$-matrix models have been used \cite{Au,Loiseau,Oset}, but only
the results corresponding to Ref.~\cite{Oset} are shown in this
letter\footnote{More complete results and discussion will be presented
elsewhere \cite{more}.}. The distinctive feature is the shape of the
$f_0(980)$ peak, which tunes the suppression of $X(3)$. Among the three
models of Refs.~\cite{Au,Loiseau,Oset}, the one corresponding to
Fig.~\ref{x2x3oset}, Ref.~\cite{Oset}, leads to the least pronounced effect. The
two other models \cite{Au,Loiseau} lead to a higher $f_0(980)$ peak, a
larger value of $\Pi(0)$ and a smaller value of $X(3)$ than shown in
Fig.~\ref{x2x3oset}: typically, at $r=25$, the upper bound for $X(3)$ would be
slightly below 0.5 and low values of $r$, such as $r\leq 15$ would even be
excluded by the stability requirement $X(3)>0$.

The sum-rule results for $X(3)$ can be converted into corresponding results
for $L_{i=4\ldots 8}$ as a function of $r$ and $F_0$ (see Fig.\ref{lecplot}).
For low $r$, the LEC's reach very large
values: their definition includes $1/B_0$ factors that make them
diverge when $X(3)\to 0$. More interesting is the enhancement of $L_5$,
$L_6$, $L_7$ and $L_8$ in the vicinity of $r\sim 25$: it reflects the drift
of $L_6(M_\rho)$ to small \emph{positive} values, which is dictated by the
sum rule. Notice that the second large-$N_c$ suppressed LEC, $L_4$, is not
predicted here: it is closely correlated to the input value of $F_0$.
Strictly speaking, $F_0$ (and $L_5$) can be constrained by studying the
slopes of the form factors in Eq.~(\ref{formfact})
\cite{Bachir1,Bachir2,Donoghue}. This procedure would yield rather low
values of $F_0$ (between 70 and 75 MeV), supporting a larger value of the
OZI-rule violating constant $L_4$ \cite{Bachir1,Bachir2,Donoghue,Meissner}.
On the other hand, the decrease of $F_0$ does not modify very much the
bounds on $X(3)$ and the consequences of Eqs.~(\ref{GOR}) and (\ref{xtwo}). For
this reason, this part of the discussion will be presented separately
\cite{more}.

{\bf 9.} A few points to summarize and conclude. {\it i)} According to the
non-perturbative formula Eq.~(\ref{GOR}), vacuum fluctuations of $\bar{q}q$
will suppress the three-flavor condensate $X(3)$, unless $L_6(M_\rho)$ is in
a narrow band around $-0.26\cdot 10^{-3}$, which further shrinks with
increasing $L_4$. The effect is nearly independent of the quark mass ratio
$r=m_s/m$. {\it ii)} Sum rules for the correlator
$\langle \bar{u}u\ \bar{s}s\rangle^c$ constrain $L_6(M_\rho)$ outside this
narrow band, suggesting that the suppression of $X(3)$ due to the vacuum
fluctuations of $\bar{q}q$ actually takes place. It is due to the importance
of the $f_0(980)$ peak ($I=J=0^{++}$) and to the QCD control of
higher-energy contributions. {\it iii)} The presence of at least three light
(massless) flavors in the sea seems crucial for this effect: vacuum
fluctuations do \emph{not} suppress the two-flavor condensate $X(2)$. The
latter remains large ($\sim 0.9-1$) unless $r<20$, see Eq.~(\ref{xtwo}).
{\it iv)} In principle, $X(2)$ can be measured in low-energy
$\pi\pi$-scattering \cite{gchipt}. If the data are sufficiently accurate,
they could be used to determine for the first time $L_8+2L_6$.
According to the present analysis, this combination of LEC's
could be (for $r=25$) 5 times larger than the values reported and used
previously \cite{daphne}. On the other hand, $\pi\pi$-scattering can hardly
tell us anything about the three-flavour condensate $X(3)$, the splitting
$X(2)-X(3)$ and about the vacuum fluctuations of $\bar{q}q$ in general.

\vspace{0.5cm}

The authors thank B.~Moussallam for many helpful discussions and for
providing his program solving numerically
Omnes-Mushkelishvili equations, and M.~Knecht, E.~Oset, H.~Sazdjian and P.~Talavera
for various discussions and comments.
Work partly supported by the EU, TMR-CT98-0169, EURODA$\Phi$NE network.

\end{document}